\documentclass[fleqn,10pt]{wlscirep}
\usepackage[utf8]{inputenc}
\usepackage[T1]{fontenc}
\usepackage{amsmath}
\usepackage{amssymb}
\usepackage{url}
\usepackage{booktabs}
\usepackage[mathscr]{euscript}
\usepackage{setspace}

\usepackage{xcolor}
\newcommand{\tcolor}{black} % revision_1
\newcommand{\tcolornew}{black} % revision_2
\newcommand{\tcolornewnew}{black} % revision_3

\newcounter{firstbib}

\title{\textcolor{\tcolor}{Global Voxel Transformer Networks for Augmented Microscopy}}

\author[1,+]{Zhengyang Wang}
\author[1,+]{Yaochen Xie}
\author[1,*]{Shuiwang Ji}
\affil[1]{Texas A\&M University, Department of Computer Science and Engineering, College Station, 77843, USA}
\affil[*]{sji@tamu.edu}
\affil[+]{These authors contributed equally to this work, and their names are listed alphabetically.}

\begin{abstract}
Advances in deep learning have led to remarkable success in augmented microscopy, enabling us to obtain high-quality microscope images without using expensive microscopy hardware and sample preparation techniques. Current deep learning models for augmented microscopy are mostly U-Net based neural networks, thus sharing certain drawbacks that limit the performance.
\textcolor{\tcolornewnew}{In particular, U-Nets are composed of local operators only and lack the ability of dynamic non-local information aggregation.}
In this work, we introduce global voxel transformer networks (GVTNets), a deep learning tool for augmented microscopy that overcomes intrinsic limitations of the current U-Net based models and achieves improved performance. GVTNets are built on global voxel transformer operators (GVTOs), which are able to aggregate global information, as opposed to local operators like convolutions. We apply the proposed methods on existing datasets for three different augmented microscopy tasks under various settings.
\end{abstract}

\begin{document}
\flushbottom
\maketitle
\doublespacing
\section{Introduction}

% 1. introduce AM and importance in bio

In modern biology and life science, augmented microscopy attempts to improve the quality of microscope images to extract more information, such as introducing fluorescent labels, increasing the signal-to-noise ratio~(SNR), and performing super-resolution. Previous advances in microscopy have allowed the imaging of biological processes with higher and higher quality~\cite{gustafsson2000surpassing,huisken2004optical,betzig2006imaging,rust2006sub,heintzmann2009subdiffraction,tomer2012quantitative,chen2014lattice,belthangady2019applications}. However, these advanced augmented-microscopy techniques usually lead to high costs in terms of the microscopy hardware and experimental conditions, resulting in many practical limitations. In addition, specific concerns are raised when recording processes of live cells, tissues, and organisms; those are, the imaging process should neither significantly affect the biological processes nor substantially harm the sample's health. For example, assessing phototoxicity is a major problem in live fluorescence imaging~\cite{laissue2017assessing,icha2017phototoxicity}. With these restrictions, high-quality microscope images are hard, expensive, and slow to obtain. While some microscope images, like transmitted-light images~\cite{selinummi2009bright}, can be collected at relatively low cost, they are not sufficient to provide accurate statistics and correct insights without augmentation. As a result, modern biologists and life scientists usually have to deal with the trade-offs between the quality of microscope images and the restrictions in the process of collecting them~\cite{pawley2006fundamental,scherf2015smart,skylaki2016challenges}.

% 2. AM is an image transformation problem in ML/DL

In recent years, the development of deep learning~\cite{lecun1998gradient} has pushed the boundaries of such trade-offs by enabling fast and inexpensive microscopy augmentation using computational approaches~\cite{sullivan2018seeing,chen2019augmented,moen2019deep}. The augmented microscopy task is formulated as a biological image transformation problem in deep learning. Specifically, models composed of multi-layer artificial neural networks take low-quality microscope images as inputs, and transform them into high-quality ones through computational processes. Deep learning has led to success in various augmented microscopy applications, such as prediction of fluorescence signals from transmitted-light images~\cite{johnson2017building,ounkomol2017three,osokin2017gans,Yuan:bioinfo18,johnson2018studying,christiansen2018silico,ounkomol2018label,belthangady2019applications}, virtual refocusing of fluorescence images~\cite{wu2019three}, content-aware image restoration~\cite{weigert2018content}, fluorescence image super-resolution~\cite{wang2018deep,wang2019deep}, and axial under-sampling mitigation~\cite{rivenson2017deep}. 

% 3. U-Net is used for image transformation problems.

Among these successful applications of deep learning, U-Net based neural networks have been the mainstream models. The U-Net was first proposed for 2D electron microscopy image segmentation~\cite{ronneberger2015u} and later extended to other biological image transformation tasks, including cell detection and quantification~\cite{falk2019u}. In the field of augmented microscopy, most deep learning models directly apply U-Net based neural networks by only changing the loss functions for training~\cite{osokin2017gans,ounkomol2018label,weigert2018content,Yuan:bioinfo18,johnson2018studying}. In general, the U-Net is an encoder-decoder framework of neural network architectures for image transformation. It consists of a down-sampling path to capture multi-scale and multi-resolution contextual information, and a corresponding up-sampling path to enable precise voxel-wise predictions. Recent studies have enhanced the U-Net by incorporating residual blocks~\cite{he2016deep,he2016identity,fakhry2017residual,lee2017superhuman} and supporting 3D image transformation~\cite{cciccek20163d}.

\begin{figure}[!t]
	\centering
	\includegraphics[width=0.9\textwidth]{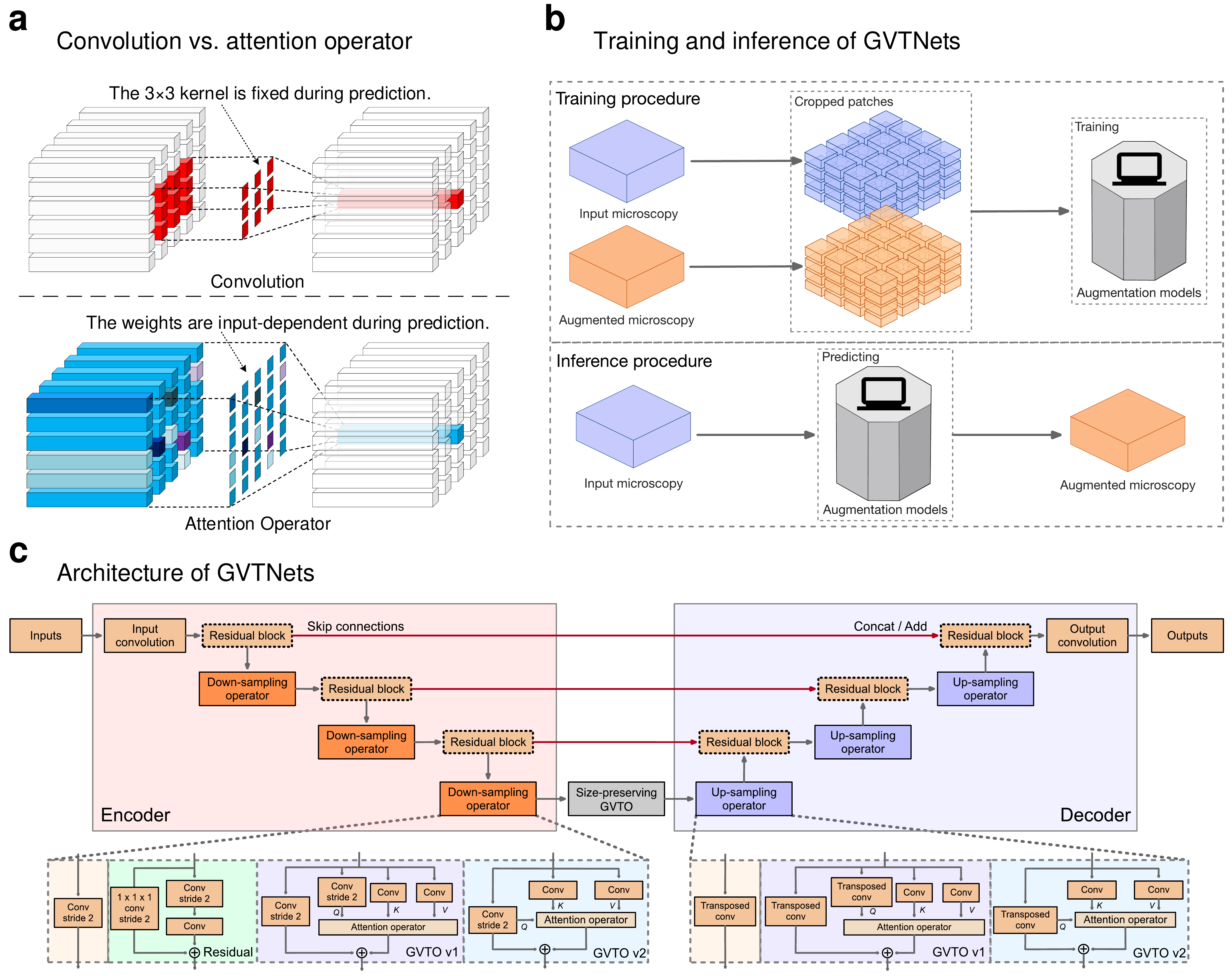}
	\caption{\textbf{GVTNets architecture, training and inference.} \textbf{a}, Comparisons between a $3 \times 3$ convolution and the attention operator in terms of receptive field and working mechanism during the inference procedure. The convolution, a typical local operator, has a fixed-size receptive field and fixed weights after training. On the contrary, the attention operator always allows a global receptive field and input-dependent weights during prediction. GVTOs, the key components of GVTNets, are built upon the attention operator. \textbf{b}, During the training procedure, registered pairs of microscope image before and after augmentation are collected and cut into small patches. During the inference procedure, the entire image is fed into the model to obtain the augmented output. \textbf{c}, A GVTNet of depth 4. The use of GVTOs differs GVTNet from the U-Net. The GVTNet fixes one size-preserving GVTO at the bottom level and allows optional GVTOs as down-sampling and up-sampling operators. The detailed description of GVTNets and GVTOs are provided in the Methods.}
	\label{fig:overall}
\end{figure}

% 4. U-Net has limitations

Despite the success of these U-Net based neural networks for augmented microscopy, we observe three intrinsic limitations caused by the fact that they implement the encoder-decoder path by stacking local operators like convolutions and transposed convolutions with small kernels. First, in local operators, the size of receptive field~(RF) of an output unit, determined by the kernel size, is usually small and does not aggregate information from the entire input (Fig.~\ref{fig:overall}a). While stacking these local operators increases the size of RF for the final output units~\cite{simonyan2014very}, the size of RF is still fixed given a specific neural network architecture. Each output unit follows a local path through the network and only has access to the information within its RF on the input image. Given a large input image, the network has to go deeper with more down-sampling and up-sampling operators to ensure each output unit received information from the entire input image. Such an approach is not efficient in terms of the amount of training parameters and computational expenses. In addition, the local path tends to focus on local dependencies among units and fails to capture long-range dependencies~\cite{vaswani2017attention,wang2018non}, which are crucial for accuracy and consistency in biological image transformation. Second, the fixed-size RF limits the model's inference performance as well. The U-Net is usually trained with small patches of paired images (Fig.~\ref{fig:overall}b), where cutting large images into small patches increases the amount of training data and stabilizes the training process by allowing large batch sizes~\cite{wilson2003general}. As the U-Net produces the output of the same spatial size as the input, it is common to feed in the entire image or patches of much larger spatial sizes than the training patches during the prediction procedure (Fig.~\ref{fig:overall}b, Supplementary Fig. 1), in order to speed up the inference~\cite{ronneberger2015u,ounkomol2018label,weigert2018content,falk2019u}. However, with the fixed-size RF, the model fails to take advantage of the knowledge from the entire input if the spatial size of the input is larger than that of RF, preventing potential inference performance boost. Third, all the local operators work with kernels whose weights are fixed after the training process, which means the importance of an input unit to an output unit is determined and not input-dependent during the inference stage (Fig.~\ref{fig:overall}a). This property is helpful in detecting and extracting local patterns~\cite{lecun1998gradient}. However, the model is supposed to be able to selectively use or ignore extracted information when transforming different input images, raising the need of operators that support input-dependent weights.

% 5. we develop new method to address the limitation

% In this work, we propose global voxel transformer networks~(GVTNets) (Fig.~\ref{fig:overall}c, Methods), an advanced deep learning tool for augmented microscopy, in order to address the limitations and improve current U-Net based neural networks. In particular, we build a family of non-local operators upon the attention operator~\cite{vaswani2017attention}, namely global voxel transformer operators~(GVTOs). In GVTOs, all the information from the input is accessible to every output unit, including both local and long-range dependencies (Fig.~\ref{fig:overall}a). In addition, the information is filtered by weights that are input-dependent, allowing a dynamic information aggregation mechanism. Therefore, GVTOs naturally overcome the above limitations of the U-Net. We develop GVTOs to support size-preserving, down-sampling and up-sampling tensor processing, which covers all kinds of operators in the U-Net framework. With GVTOs, we build our GVTNets based on the U-Net framework. Specifically, GVTNets connect the down-sampling and up-sampling paths using the size-preserving GVTO at the bottom level, and allow users to flexibly use more GVTOs to replace local operators in the U-Net. It is worth noting that, while GVTNets follow the U-Net framework, GVTOs can be used in other kinds of networks as well.

\textcolor{\tcolor}{In this work, we argue that all three limitations above can be addressed by introducing the attention operator~\cite{vaswani2017attention} into U-Net based neural networks. In order to demonstrate this point, we compare the attention operator with a typical local operator, \emph{i.e.}, convolution (Fig.~\ref{fig:overall}a). There are essential differences between the convolution and the attention operator. On one hand, the convolution has a local RF determined by its kernel, where each output unit receives information from a local area of input units. Meanwhile, note that the kernel weights are fixed after training. In other words, the weights do not depend on inputs during the inference. On the other hand, the attention operator computes each output unit as a weighted sum of all input units, where the weights are obtained through interactions between different representations of the inputs (Methods). As a result, the attention operator is a non-local operator with a global receptive field, which can potentially overcome the first two limitations. In addition, the weights in the attention operator are input-dependent, addressing the third limitation.}

\textcolor{\tcolor}{Based on this insight, we build a family of non-local operators upon the attention operator, namely global voxel transformer operators~(GVTOs) (Fig.~\ref{fig:overall}c, Methods). GVTOs organically combine local and non-local operators (Supplementary Fig. 11) and can capture both local and long-range dependencies. In particular, GVTOs extend the attention operator to serve as flexible building blocks in the U-Net framework. Specifically, we develop GVTOs to support not only size-preserving, but also down-sampling and up-sampling tensor processing, which covers all kinds of operators in the U-Net framework. It is worth noting that, while GVTOs are designed for the U-Net framework, they can also be used in other kinds of networks as well.}

\textcolor{\tcolor}{With GVTOs, we propose global voxel transformer networks~(GVTNets) (Fig.~\ref{fig:overall}c, Methods), an advanced deep learning tool for augmented microscopy, in order to address the limitations and improve current U-Net based neural networks. GVTNets follow the same encoder-decoder framework as the U-Net while using GVTOs instead of local operators only. To be concrete, we force GVTNets to connect the down-sampling and up-sampling paths using the size-preserving GVTO at the bottom level, which separates GVTNets from the U-Net. In addition, we allow users to flexibly use more GVTOs to replace local operators in the U-Net framework.}

% 6. summary of results

In the following, we (1) demonstrate the power of the basic GVTNets where only one size-preserving GVTO at the bottom level is applied, (2) show the effectiveness of employing more GVTOs in GVTNets and point out how GVTNets improve the inference performance, (3) explore the use of GVTOs in more complex and composite models, and (4) investigate the generalization ability of GVTNets. All the experiments are conducted on publicly available datasets for augmented microscopy~\cite{ounkomol2018label,weigert2018content,moen2019deep}.

\section{Results}

% A figure showing the training/inference procedures and the overall architecture should be included.

\subsection{Global voxel transformer networks training and inference}

% Explain training patch, prediction patch, how to train and inference briefly.

Global voxel transformer networks~(GVTNets) are trained end-to-end under a supervised learning setting through back-propagation~\cite{lecun1998gradient} (Methods). While the model aims at augmenting microscopy computationally, it still requires a relatively small amount of augmented microscopy images to be collected for training. Specifically, the training data are registered pairs of biological images before and after augmentation. Once trained, the model can be used to augment microscope images \textit{in silico}, without involving any expensive microscopy hardware and technique. Following previous studies, we crop the training images into patches of smaller spatial sizes to train GVTNets. However, during the inference procedure, we feed in the entire image for prediction (Fig.~\ref{fig:overall}b). Note that GVTNets are able to handle inputs of any spatial size, and in particular, tend to perform better given inputs of larger spatial sizes due to the ability of utilizing global information from the entire input. The power of GVTNets come from the use of global voxel transformer operators~(GVTOs), which are inherently different from local operators as well as the fully-connected layers in deep learning (Methods).

\begin{figure}[!t]
	\centering
	\includegraphics[width=0.9\textwidth]{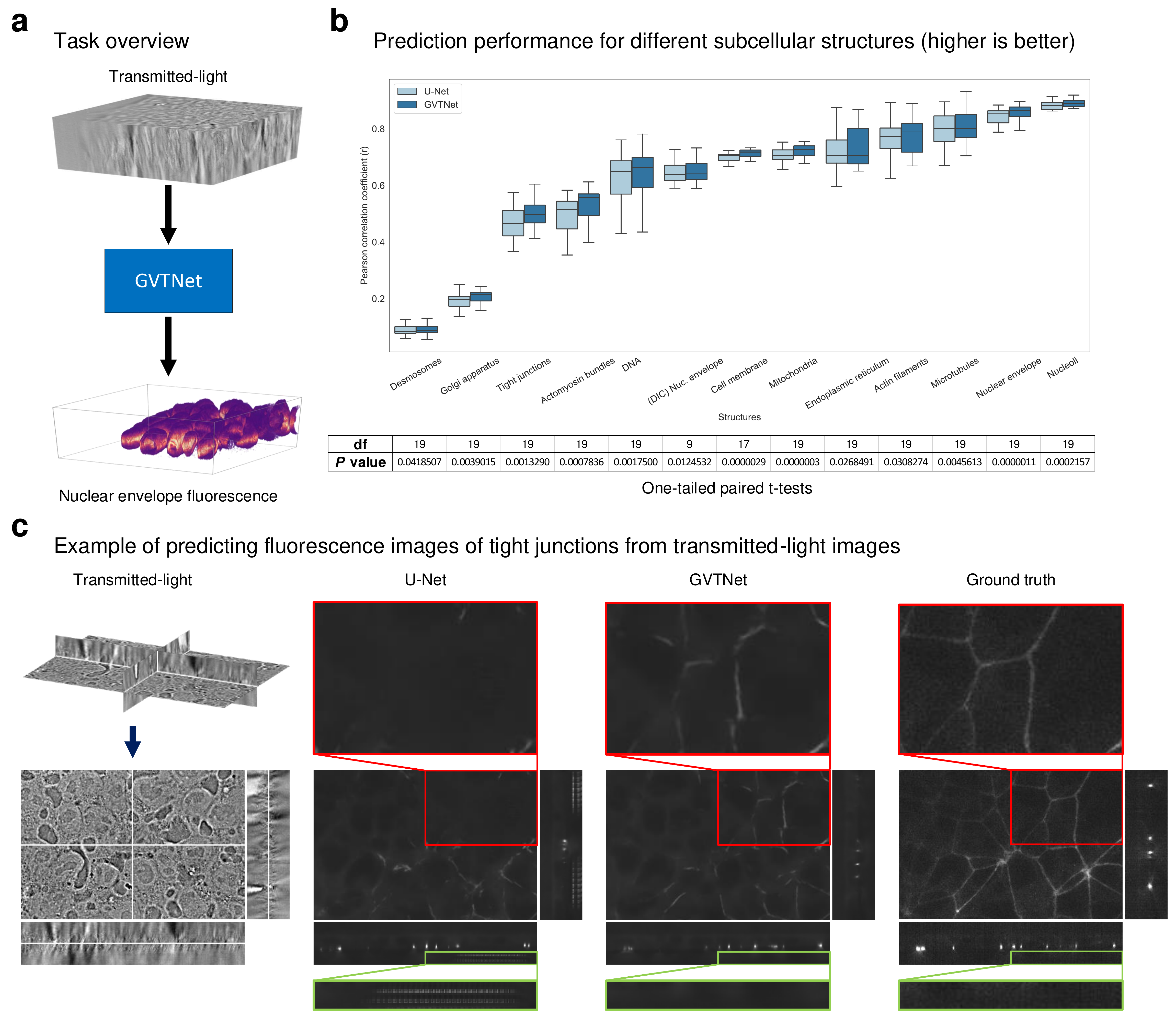}
	\caption{\textbf{GVTNets on label-free prediction of 3D fluorescence images from transmitted-light microscopy.} \textbf{a}, This augmented microscopy task is to predict the fluorescence images of subcellular structures from inexpensive transmitted-light images without fluorescent labels. \textbf{b}, Top: Distributions of the Pearson correlation coefficient~(r) between the ground truth images and the predicted images of the GVTNet on the testing datasets for 13 different subcellular structures. Each pair of images leads to a point in the distribution. In the box and whisker plots, the 25th, 50th, and 75th percentile points are marked by the box, and whiskers indicate the minimum and maximum. The number of testing images is 18 for the cell membrane, 10 for the differential interference contrast~(DIC) nuclear envelope, and 20 for the others. Bottom: One-tailed paired t-test results on the performance of the GVTNet and the U-Net baseline. The degree of freedom is the number of testing images minus one. All the \textit{P} values are smaller than 0.05. \textbf{c}, From left to right, columns are transmitted-light input images, the predicted fluorescence images using the U-Net baseline, the predicted fluorescence images using the GVTNet, and the ground truth fluorescence images. We visualize the center z-, y-, and x-slices for 3D images. Clearly, the GVTNet captures more details. In addition, the GVTNet avoids artifacts caused by local operators.}
	\label{fig:label_free}
\end{figure}

\subsection{Label-free prediction of 3D fluorescence images from transmitted-light microscopy}

% demonstrate the power of the basic GVTNets where only one size-preserving GVTO at the bottom level is applied

We first ask whether basic GVTNets achieve improved performance over U-Net based neural networks. A basic GVTNet differs from the U-Net only at the bottom level, by using a size-preserving GVTO instead of convolutions. The replacement is crucial, giving each output unit access to information from the entire input image, regardless of the spatial size. We apply a basic GVTNet on the public dataset from C. Ounkomol et al.~\cite{ounkomol2018label}, where the task is label-free prediction of 3D fluorescence images from transmitted-light microscopy (Fig.~\ref{fig:label_free}a).

The dataset is composed of 13 datasets corresponding to 13 different subcellular structures.
\textcolor{\tcolor}{All the images in the datasets are spatially registered and obtained from a database of images produced by the Allen Institute for Cell Science’s microscopy pipeline~\cite{ounkomol2018label}. The training and testing splits are provided by C. Ounkomol et al.~\cite{ounkomol2018label} and available in our published code.}
For each structure, the training data are 30 spatially registered pairs of 3D transmitted-light images and ground truth fluorescence images. The number of testing images is 18 for the cell membrane, 10 for the differential interference contrast~(DIC) nuclear envelope, and 20 for the others.

\textcolor{\tcolor}{We use the model proposed by C. Ounkomol et al.~\cite{ounkomol2018label} as the baseline model, which is the current state-of-the-art model on the 13 datasets.}
The baseline model is a U-Net based neural network of depth 5 containing $23,280,769$ training parameters, while the basic GVTNet that we used is of depth 4 containing $6,172,225$ training parameters (Supplementary Fig. 2). As a result, the basic GVTNet has only $26.5\%$ of training parameters of the baseline model. In addition, the computation speed becomes faster; that is, the GVTNet takes $0.4$s to make prediction for one 3D image while the U-Net takes $1$s~\cite{ounkomol2018label}.

We quantify the model performance by computing the Pearson correlation coefficient on the testing data (Methods).
On all of the 13 datasets, our basic GVTNet consistently outperforms the U-Net baseline. We perform one-tailed paired t-tests and obtain \textit{P} values smaller than 0.05 for all datasets, showing the improvements are statistically significant (Fig.~\ref{fig:label_free}b).
The visualization of predictions indicates that the GVTNet captures more details than the U-Net baseline due to the access to more information, and is able to use global information to avoid local inconsistency (Fig.~\ref{fig:label_free}c).
The quantitative testing results in terms of Pearson correlation coefficients are provided in Supplementary Table 1.
Examples of predictions on testing images for all 13 structures can be found in Supplementary Fig. 3.
These experimental results indicate the effectiveness of only one size-preserving GVTO and the resulted basic GVTNets.

\textcolor{\tcolornewnew}{We note that both GVTNets and the U-Net baselines perform poorly on the datasets corresponding to Golgi apparatus and desmosomes subcellular structures. According to C. Ounkomol et al.~\cite{ounkomol2018label}, a possible explanation is that the correlations between the input transmitted-light microscope images and the target fluorescence images are weak in these two datasets. As most supervised deep learning methods models try to capture the correlations between inputs and outputs during training, the inference performance could be poor if the correlations are weak. Nevertheless, a global view is helpful, as more correlations are considered when making the prediction for each voxel. And GVTNets indeed improve the performance on these datasets. In addition, we also find that the amount of improvements brought by GVTNets vary for different datasets. We suspect that this is due to the characteristics of different subcellular structures. For example, the shapes are sparser for actomyosin bundles and tight junctions (Supplementary Fig. 3). It may be more important to capture the correlations among distant voxels when predicting these subcellular structures, where GVTNets achieve more improvements.}

\begin{figure}[!t]
	\centering
	\includegraphics[width=0.9\textwidth]{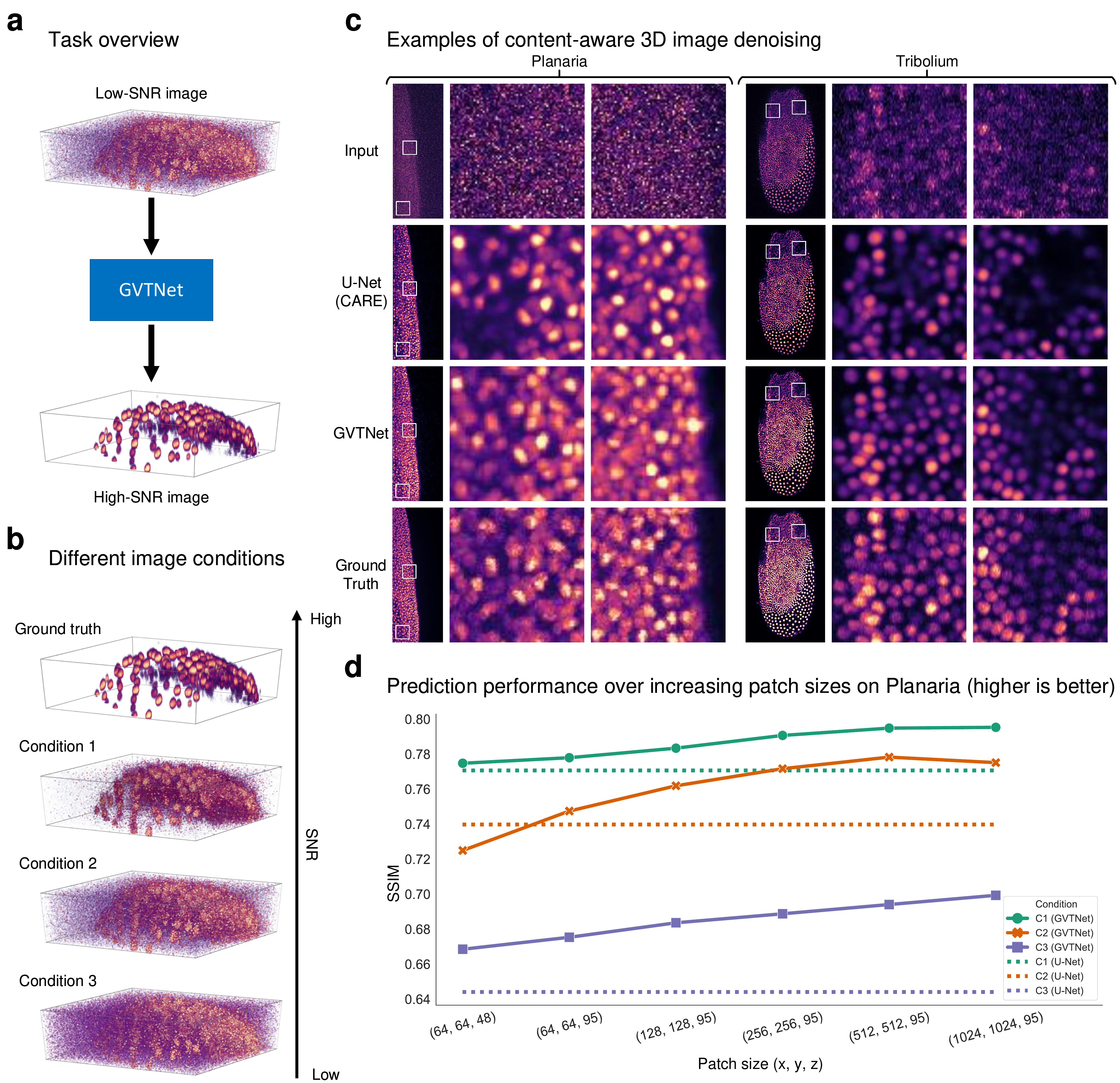}
	\caption{\textbf{GVTNets on content-aware 3D image denoising.} \textbf{a}, This augmented microscopy task is to improve the SNR by removing the noises from the low-SNR images captured in poor imaging conditions. \textbf{b}, The ground truth image captured with full exposure and lazer power condition, along with three noised images captured in weaker conditions at three different levels (C1-C3). \textbf{c}, From top to bottom, rows are the input noisy images, the predicted denoised images using the U-Net based CARE, the predicted denoised images using the GVTNet, and the ground truth denoised images. On both \textit{Planaria} and \textit{Tribolium} datasets, the U-Net fails to capture details in input regions with weak signals. On the contrary, the GVTNet obtains more precise predictions with more details in such regions. \textbf{d}, The inference performance in terms of SSIM over increasing prediction patch sizes on the \textit{Planaria} dataset. The number of testing images is 20. Dotted lines represent the U-Net and solid lines represent the GVTNet. The inference performance of the GVTNet increases with larger prediction patch sizes, showing its ability of utilizing knowledge from the entire input.}
	\label{fig:denoising}
\end{figure}

\subsection{Content-aware 3D image denoising}

% show the effectiveness of employing more GVTOs in GVTNets

Next, we explore the potential of GVTNets by applying more GVTOs. Specifically, we apply GVTNets with both size-preserving and up-sampling GVTOs on two independent content-aware 3D image denoising tasks (Fig.~\ref{fig:denoising}a); namely, improving the signal-to-noise ratio~(SNR) of live-cell imaging of \textit{Planaria S. mediterranea} and developing \textit{Tribolium castaneum} embryos.

The datasets were published by M. Weigert et al.~\cite{weigert2018content}, which contain pairs of 3D low-SNR images and ground truth high-SNR images for training and testing. The training data are provided in the form of 17,005 and 14,725 small cropped patches of size $64\times 64\times 16$ for \textit{Planaria} and \textit{Tribolium} datasets, while the testing data are 20 testing images of size $1024\times1024\times95$ and 6 testing images of average size around $700\times 700\times 45$ for the two datasets, respectively. In addition, the testing data come with three image conditions referring to three different SNR levels, leading to three degrees of denoising difficulty (Fig.~\ref{fig:denoising}b).
\textcolor{\tcolor}{Here, the image conditions refer to the laser-power and exposure-time during image collection~\cite{weigert2018content}. Generally, low laser-power and short exposure-time lead to low SNR levels. Concretely, in the \textit{Planaria} dataset, four different laser-power/exposure-time conditions are used: GT (ground truth) and C1–C3, specifically 2.31 mW/30 ms (GT), 0.12 mW/20 ms (C1), 0.12 mW/10 ms (C2), and 0.05 mW/10 ms (C3). Similarly, in the \textit{Tribolium} dataset, four different laser-power imaging conditions are used: GT and C1–C3, specifically 20 mW (GT), 0.5 mW (C1), 0.2 mW (C2), and 0.1 mW (C3).}
As a result, each ground truth high-SNR image in testing dataset has three corresponding low-SNR images.

The baseline models in these experiments are the content-aware image restoration~(CARE) networks~\cite{weigert2018content}, which are based on the 3D U-Net~\cite{cciccek20163d}.
\textcolor{\tcolor}{The U-Net based CARE networks achieve the current best performance on these two datasets, serving as a strong baseline.} We build a GVTNet by replacing the bottom convolutions and up-sampling operators with corresponding size-preserving and up-sampling GVTOs (Supplementary Fig. 4).

In order to quantify the model performance, we compute two evaluation metrics, \textit{i.e.}, the structural similarity index~(SSIM)~\cite{wang2004image} and normalized root-mean-square error~(NRMSE) (Methods). The models are evaluated under three SNR levels individually.
The visualization results demonstrate that the GVTNet can take advantage of long-range dependencies to recover more details in areas with weak signals than the U-Net (Fig.~\ref{fig:denoising}c).
The quantitative results also indicate significant and consistent improvements of the GVTNet over the U-Net based CARE under all image conditions on both datasets, revealing the advantages of GVTNets with more GVTOs (Supplementary Fig. 5, Supplementary Table 2). 
More examples of predictions on testing images can be found in Supplementary Fig. 6-7.

% provide insights on how GVTNets improve the performance

\textcolor{\tcolornewnew}{In order to provide insights on how GVTNets improve the inference performance by utilizing global information, we conduct extra experiments by varying the spatial sizes of input images during the inference process. To be specific, as both GVTNets and the U-Net are able to handle inputs of any spatial size, we can either feed the entire image directly into the model or crop the image into small prediction patches and reconstruct the entire augmented image after prediction (Supplementary Fig. 1). Theoretically, since the size of receptive filed~(RF) in the U-Net is fixed and local, the prediction results will be the same as long as the size of prediction patches is larger than that of RF. On the other hand, the size of RF in GVTNets is dynamic and global and always covers the entire input image. This property allows the use of more knowledge for better inference performance given large prediction patches, even if the training patches have much smaller sizes. In order to verify this insight, we train the GVTNet and CARE on the \textit{Planaria} dataset and compare prediction results in terms of SSIM when using prediction patches of sizes ranging from $64\times 64\times 48$ to $1024\times 1024\times 95$ (entire image size). The results are summarized in Fig.~\ref{fig:denoising}d. Examples of predictions are provided in Supplementary Fig. 8 and detailed performance comparisons with error bars are provided in Supplementary Fig. 9. The prediction results of the U-Net remain the same when increasing prediction patch sizes, forming a horizontal line (Supplementary Fig. 1). On the contrary, significant improvements can be observed for the GVTNet. These results show a unique advantage of GVTNets that a performance boost can be achieved by using larger prediction patches. This advantage can be easily achieved without the need to retrain the model, which is expensive and time-consuming. However, note that larger prediction patch sizes do not always lead to better performance. The correlations among voxels will get weaker and weaker with the increasing distance. It is possible that more noises are aggregated than useful information when the prediction patch size reaches a certain point. For example, in Fig.~\ref{fig:denoising}d, we can observe a local maximum when the prediction patch size is $512\times 512\times 95$ for C2.}

\textcolor{\tcolornewnew}{}

\begin{figure}[!t]
	\centering
	\includegraphics[width=0.9\textwidth]{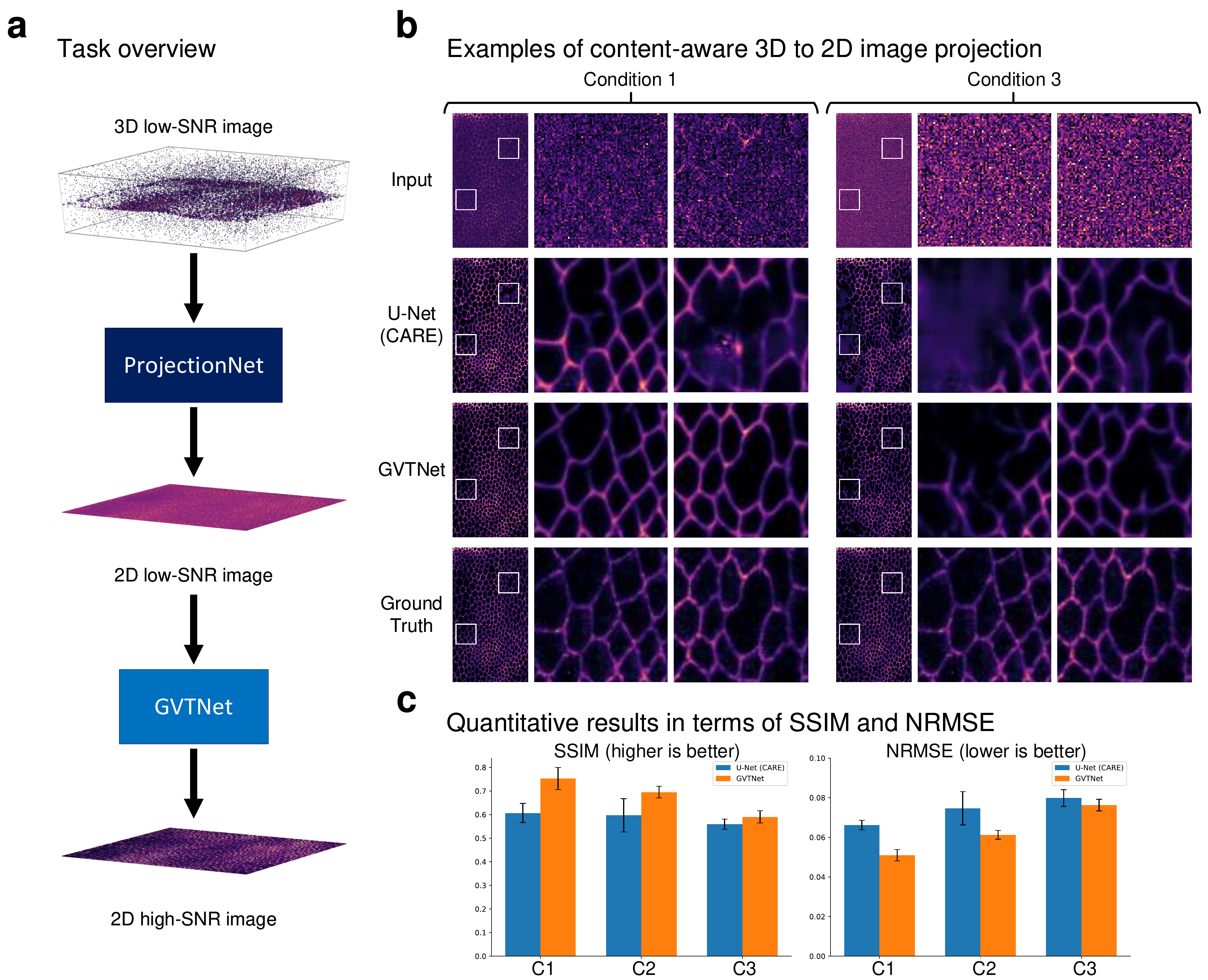}
	\caption{\textbf{GVTNets on content-aware 3D to 2D image projection.} \textbf{a}, This augmented microscopy task is to project a low-SNR 3D image into a high-SNR 2D surface. The model consists of a ProjectionNet that produces an intermediate 2D low-SNR image and a 2D GVTNet that outputs the high-SNR 2D image. \textbf{b}, From top to bottom, rows are input images, the predicted images using the U-Net based CARE, the predicted images using the GVTNet, and the ground truth images. Visualization results indicate that the U-Net is more sensitive to the irregular input voxel values and collapses in surrounding areas. In addition, the U-Net tends to give ambiguous and blurred predictions where the input information is insufficient. On the contrary, the GVTNet is more robust to these cases. \textbf{c}, Prediction performance on the testing data of the \textit{Flywing} dataset, in terms of SSIM and NRMSE under three imaging conditions. The number of testing images is 26. The 68$\%$ confidence intervals are marked by computing the standard deviation over testing images.}
	\label{fig:flywing}
\end{figure}

\subsection{Content-aware 3D to 2D image projection}

% explore the use of GVTOs in more complex and composite models

While we use GVTOs to build GVTNets, GVTOs are a family of operators that support any size-preserving, down-sampling and up-sampling tensor processing and can be used outside GVTNets. Therefore, we further examine the proposed GVTOs on more complicated and composite models. In particular, we apply GVTOs and GVTNets on the 3D \textit{Drosophila melanogaster Flywing} surface projection task~\cite{aigouy2010cell, etournay2015interplay} (Fig.~\ref{fig:flywing}a).

The model for this task is supposed to take a noised 3D image as the input and projects it into a denoised 2D surface image. The typical deep learning model involves two parts; those are, a network for 3D to 2D surface projection, followed by a network for 2D image denoising. For example, \textcolor{\tcolor}{the current best model}, CARE~\cite{weigert2018content}, uses a task-specific convolutional neural network~(CNN)~\cite{lecun1998gradient} for projection and a 2D U-Net for denoising. The task-specific CNN is also composed of convolutions, down-sampling and up-sampling operators. We design our model based on CARE by applying GVTOs in the first CNN and replace the 2D U-Net with a 2D GVTNet (Supplementary Fig. 10). The resulted composite model employs size-preserving and up-sampling GVTOs in both parts.

We compare our model with CARE on the \textit{Flywing} dataset~\cite{weigert2018content} in terms of SSIM and NRMSE. The dataset contains 16,891 pairs of small 3D noisy image patches and ground truth 2D surface image patches for training, and 26 complete images for testing.

The quantitative results indicate that the composite model augmented by GVTOs achieves significant improvements (Fig.~\ref{fig:flywing}c). We provide the detailed quantitative results in Supplementary Table 3. The visualization results show that the GVTOs have a stronger capability to recognize non-noisy objects at regions of lower SNR within an image, where the original model tends to fail (Fig.~\ref{fig:flywing}b). This is because the global information is of great importance to the projection tasks, especially along the Z-axis, \textcolor{\tcolor}{where the projection happens. Specifically, for each (x, y) location in the 3D image, only one voxel along the Z-axis will be projected to the 2D surface. This restriction is only available when the model has the global information along the Z-axis.}
Therefore, plugging GVTOs into the projection process can effectively improve the overall performance. More examples of predictions on testing images can be found in Supplementary Fig. 11.

\begin{figure}[!t]
	\centering
	\includegraphics[width=0.9\textwidth]{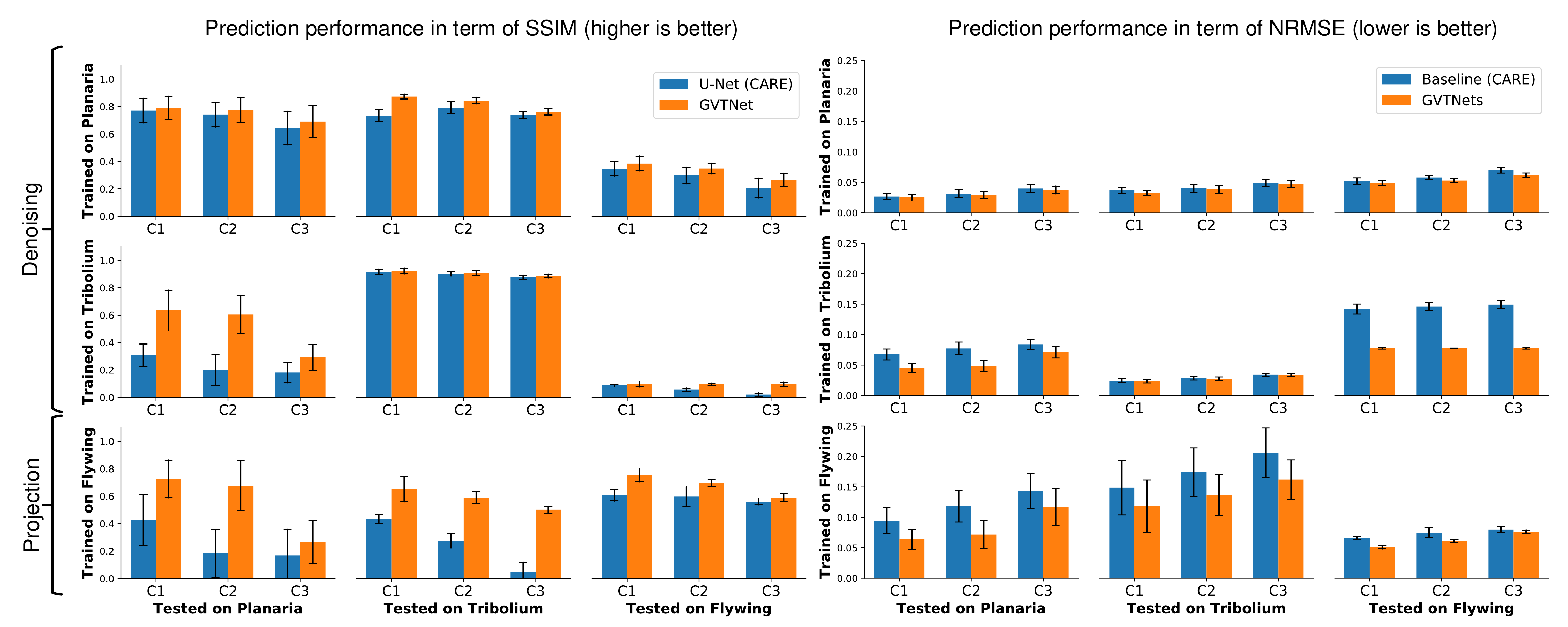}
	\caption{\textbf{Generalization ability of GVTNets.} Comparisons of the transfer learning performance between the U-Net base CARE and our GVTNets, in terms of SSIM (left) and NRMSE (right). Rows represent the dataset on which the models are trained and columns represent the dataset on which the models are tested. The first two rows correspond to the 3D denoising tasks and the third row corresponds to the 3D to 2D projection tasks. The diagonal charts are the performance of models trained and tested on the same datasets. The GVTNets can achieve a promising performance under this simplest transfer learning setting, due to the input-dependent weights of GVTOs. The 68$\%$ confidence intervals are marked by computing the standard deviation over testing images.}
	\label{fig:transfer}
\end{figure}

\subsection{Transfer learning ability of GVTNets}

% investigate the generalization ability of GVTNets

We have shown the effectiveness of GVTNets for augmented microscopy applications under a supervised learning setting. In the following, we further investigate the generalization ability of GVTNets under a simple transfer learning setting~\cite{pan2009survey}, where we train GVTNets on one dataset and perform testing on other datasets for the same task.
In this case, the inconsistencies between the training and testing data often lead to the collapse of models based on local operators, such as the U-Net. One reasonable explanation is that the weights of kernels in local operators are fixed after training and independent to the inputs~\cite{pan2009survey}. This limits the ability to deal with the different data distributions in training and inference procedures.

As GVTOs achieve input-dependent weights, we hypothesize that GVTNets are more robust to such inconsistencies and have a better generalization ability. We conduct experiments to verify the hypothesis using the three datasets from M. Weigert et al.~\cite{weigert2018content}; namely, the \textit{Planaria}, \textit{Tribolium} and \textit{Flywing} datasets. Note that all these datasets originally have 3D high-SNR ground truth images for the 3D denoising task. By applying PreMosa~\cite{blasse2017premosa} on the 3D ground truth images, we can obtain 2D ground truth images for the 3D to 2D projection task. Therefore, these datasets can be used in either task for both training and testing. The baseline models are still the U-Net based CARE networks in these experiments, and we use the same GVTNet as introduced above for comparison (Supplementary Fig. 4, Supplementary Fig. 10). In general, we train GVTNet and CARE on one of the three datasets, and compare their testing performance on the remaining two datasets, resulting in three sets of experiments. To be concrete, the first two experiments where either the \textit{Planaria} or \textit{Tribolium} dataset is used for training are doing the 3D denoising tasks. The third experiment where models are trained on the \textit{Flywing} dataset is performing the 3D to 2D projection task.

The comparison results in terms of SSIM and NRMSE are shown in Fig.~\ref{fig:transfer}. The detailed quantitative results can be found in Supplementary Table 4. GVTNet obtains a more promising transfer learning performance than CARE, indicating a better generalization ability.

\section{Discussion}

% Summarize our contributions.

We have introduced GVTNets built on GVTOs, an advanced deep learning tool for augmented microscopy. Compared to the U-Net, GVTNets are more powerful models that are capable of capturing long-range dependencies and selectively aggregating global information for inputs of any spatial size. With GVTNets, various augmented microscopy tasks can be performed with significantly improved accuracy, such as predicting the fluorescence images of subcellular structures directly from transmitted-light images without using fluorescent labels, conducting content-aware image denoising, and projecting a 3D microscope image to a 2D surface for analysis. We have demonstrated the superiority of GVTNets and GVTOs on several publicly available datasets for augmented microscopy~\cite{ounkomol2018label,weigert2018content,moen2019deep}. In particular, we have provided examples where GVTNets achieve better inference performance with inputs of larger spatial sizes, indicating the ability of utilizing global information. In addition, besides the supervised learning setting, GVTNets outperform the U-Net under a simple transfer learning setting, showing better generalization ability due to input-dependent weights.

% impact

\textcolor{\tcolornew}{We anticipate that our work would exert potential impacts on biological image analysis in general and augmented microscopy specifically. Image analysis plays an indispensable role in biological research, where machine learning methods and tools have been widely used and dramatically advanced biological research and discoveries. In particular, the past decade has witnessed revolutionary changes in machine learning with the rapid developments of deep learning~\cite{lecun1998gradient}. Recent studies~\cite{ounkomol2018label,christiansen2018silico,Yuan:bioinfo18,wang2018deep,weigert2018content,Cai:KDD18} have shown that deep learning allows biological research to transcend the limits imposed by imaging hardware, enabling discoveries at scales and resolutions that were previously impossible. We observe that most of these biological image analysis tasks can be formulated as biological image transformation problems~\cite{moen2019deep}. In such tasks, the U-Net~\cite{ronneberger2015u,falk2019u} is the most popular and successful deep model, achieving the state-of-the-art performances~\cite{ounkomol2018label,christiansen2018silico,wang2018deep,weigert2018content,moen2019deep}. Our proposed GVTNets can be directly used to replace the U-Net and boost the performance by addressing intrinsic limitations of the U-Net. Specifically, our experimental results have shown the superiority of GVTNets in various augmented microscopy tasks. These results are expected to have an immediate and strong impact on basic biology by enabling discoveries, observations, and measurements that were previously unobtainable. In addition, since the limitations of the U-Net are general and not task-specific, we anticipate that GVTNets will improve the U-Net in other biological image transformation tasks and potentially benefit a wider range of biological research based on image analysis. Last but not least, from the practical perspective, the deployment of solution is as important as developing new solutions~\cite{moen2019deep}. To make GVTNets easy to use in various biological image transformation tasks, we publish our code as an open-source tool with detailed instructions (Supplementary Note 2). Our code may greatly benefit both biology and computer science research communities.}

% Comparison to previous studies.

\textcolor{\tcolor}{In the literature, there exist many other studies that attempt to improve the U-Net in various aspects~\cite{zhang2017image,huang2018range,oktay2018attention,zhou2018unet++,gu2019net}. Among them, some studies~\cite{zhang2017image,huang2018range,gu2019net} explore a similar direction to our work, which is to allow the U-Net to capture long-range dependencies or global context information. They can be mainly divided into two categories. One is to add modules composed of dilated convolutions, like Zhang et al.~\cite{zhang2017image} and CE-NET~\cite{gu2019net}. Dilated convolutions can expand the receptive field of convolutions to capture longer-range dependencies. However, they are still local operators in essence, sharing similar limitations. For example, they cannot collect global information when inputs become larger than the receptive field. The other category is to apply global pooling to extract global information and use it to facilitate local operators, such as RSGU-Net~\cite{huang2018range}. However, important spatial information is lost during global pooling, which potentially limits the performance. Different from these two categories, we extend the attention operator to achieve the goal. To demonstrate the advantages of our method over previous methods, we compare GVTNets with representative models, \emph{i.e.}, RSGU-Net~\cite{huang2018range} and CE-NET~\cite{gu2019net}, on content-aware 3D image denoising tasks, as reported in Supplementary Table 5. Our method outperforms both methods significantly, with similar computational cost.}

\textcolor{\tcolornew}{Other studies~\cite{oktay2018attention,zhou2018unet++} improve the U-Net in orthogonal directions. Oktay et al.~\cite{oktay2018attention} propose to add the gate mechanism to the skip connections, filtering out irrelevant information. It is worth noting that the gate mechanism and the attention mechanism are essentially different in terms of computation, functionality, and flexibility. The gate mechanism performs spatially element-wise filtering so that there is no explicit communication between spatial locations. On the contrary, the attention mechanism aggregates information from all spatial locations (Methods). Moreover, the gate mechanism can only be used for size-preserving tensor processing, while the attention mechanism can be extended for down-sampling and up-sampling tensor processing by our GVTOs. Zhou et al.~\cite{zhou2018unet++} propose a nested U-Net architecture by adding dense skip connections. The nested architecture facilitates the training and yields better inference performance.}

\textcolor{\tcolor}{In terms of augmenting images with deep learning methods, generative adversarial network~(GAN)~\cite{goodfellow2014generative} is a promising choice~\cite{Cai:KDD18,Yuan:bioinfo18,moen2019deep,rivenson2019virtual}. We point out that GAN based methods are orthogonal to our GVTNets in the sense that they can be used together. Note that GAN is composed of a generator and a discriminator. In GAN based image augmentation models, the generator is typically a U-Net~\cite{rivenson2019virtual}, which we can improve with our GVTNets. We conduct experiments on content-aware 3D image denoising tasks. The results can be found in Supplementary Table 6. As indicated by the results, under the GAN framework, our GVTNets can improve the U-Net as well.}

% Drawbacks of our proposed model. However, (why these drawbacks do not hinder our contributions) ...

The key components of GVTNets are GVTOs. One concern about GVTOs is the efficiency. Given the inputs of the same size, GVTOs usually require more time and take up more memory for computation than local operators like convolutions. This is due to the use of the self-attention operator. However, the high cost of GVTOs does not necessarily make GVTNets more expensive than the U-Net. By taking advantage of the more powerful GVTOs, the overall network architecture can be simpler, improving the efficiency. For example, in the label-free fluorescence image prediction experiments, we have shown that a GVTNet can outperform a U-Net based neural network with only $26.5\%$ of training parameters and faster computation speed.

Another limitation of GVTNets is the shared disadvantage of current deep learning models~\cite{ounkomol2018label,weigert2018content}. Models trained on one biological image transformation dataset can hardly be used for another dataset. Therefore, high-quality training data must be collected for each task, which is expensive and time-consuming. GVTNets have shown promising improvements under the simplest transfer learning setting without fine-tuning. We anticipate that the combination of GVTNets and recent advances of transfer learning~\cite{pan2009survey} and meta learning~\cite{finn2017model} can greatly alleviate this limitation.

\section*{Acknowledgements}

We thank the CARE and the Allen Institute for Cell Science teams for making their data and tools publicly available. This work was supported in part by National Science Foundation grants DBI-1922969, IIS-1908166, and IIS-1908220, National Institutes of Health grant 1R21NS102828, and Defense Advanced Research Projects Agency grant N66001-17-2-4031.

\section*{Author contributions}

Z.W. and Y.X. shared first-authorship.
S.J. conceived and initiated the research. Z.W. and S.J. designed the methods.
Z.W. and Y.X. implemented the training and validation methods.
Z.W. and Y.X. designed and developed the software package.
S.J. supervised the project.
Z.W., Y.X., and S.J wrote the manuscript.

\section*{Competing Interests}

The authors declare no competing interests.

\clearpage
	
\section{Methods}

% This section focuses on giving details for reproduction, including detailed model architectures for different tasks, hyper-parameter settings, loss functions, optimazation methods, evaluation metrics computation.... Any detail that we think is necessary can be included.

\subsection{Network architecture}

\subsubsection{General framework}\label{sec:unet_framework}

% U-Net framework

Global voxel transformer networks~(GVTNets) follow the same encoder-decoder framework as the U-Net~\cite{ronneberger2015u,cciccek20163d,falk2019u}, which represents a family of deep neural networks for biological image transformations. An encoder takes the image to be transformed as the input and computes feature maps of gradually reduced spatial sizes, which encode multi-scale and multi-resolution information from the input image. Then a corresponding decoder uses these feature maps to produce the transformed image, during which feature maps of gradually increased spatial sizes are computed. GVTNets support both 2D and 3D biological image transformations. We use the 3D case to describe the architecture in detail (Fig.~\ref{fig:overall}c).

% Encoding: first conv & down-sampling & depth

In our GVTNets, the encoder starts with an initial $3 \times 3 \times 3$ convolution that transforms the input image into a chosen number of feature maps of the same spatial size, initializing the encoding. The encoding process is achieved by down-sampling operators interleaved with optional size-preserving operators. Each down-sampling operator halves the size along each spatial dimension of feature maps but doubles the channel dimension, \textit{i.e.}, the number of feature maps. To be specific, given an $d \times h \times w \times c$ tensor representing $c$ feature maps of the spatial size $d \times h \times w$ as inputs, a down-sampling operator will output an $d/2 \times h/2 \times w/2 \times 2c$ tensor. Feature maps of the same spatial size are considered at the same level. As a result, the number of levels, also known as the depth of the network, is determined by the number of down-sampling operators in the encoder.

% Decoding: up-sampling

Correspondingly, the decoder is composed of the same number of up-sampling operators interleaved with optional size-preserving operators. The decoding process computes feature maps of increased spatial sizes in a level-by-level fashion, where each up-sampling operator doubles the size along each spatial dimension of feature maps but halves the channel dimension, as opposed to down-sampling operators. Therefore, there is a one-to-one correspondence between down-sampling and up-sampling operators. The decoder ends with an output convolution that outputs transformed image of the same spatial size as the input image.

% Connections

The encoder and decoder are connected at each level. The bottom level contains the outputs of the encoder, which are feature maps of the smallest size in the U-Net framework. These feature maps, after optional size-preserving operators, serve as inputs to the decoder. In upper levels, there exist skip connections between the encoder and decoder. Concretely, the input feature maps to each down-sampling operator are concatenated or added to the output feature maps of the corresponding up-sampling operator. The skip connections allow the decoder to take advantage of encoded multi-scale and multi-resolution information, which increases the capability of the framework and facilitates the training process~\cite{ronneberger2015u,lee2017superhuman}.

\subsubsection{Global voxel transformer networks}

% Options of operators

The major difference between our GVTNets and the original U-Net lies in the choices of the size-preserving, down-sampling, and up-sampling operators. GVTNets are equipped with global voxel transformer operators~(GVTOs), which can be flexibly used for size-preserving, down-sampling, or up-sampling tensor processing. In particular, GVTNets fix the size-preserving operator at the bottom level to be the size-preserving GVTO, ensuring that global information is encoded and aggregated before going through the decoder. The other size-preserving operators are set to pre-activation residual blocks~\cite{he2016identity}, consisting of two $3 \times 3 \times 3$ convolutions with the ReLU activation function~\cite{krizhevsky2012imagenet} (Supplementary Fig. 12a). Down-sampling and up-sampling GVTOs can be used as corresponding operators based on the datasets and tasks.

\subsection{Global voxel transformer operators}

As described above, the key components of our GVTNets are global voxel transformer operators~(GVTOs), which are able to selectively use long-range information among input units. We take the 3D case to illustrate the size-preserving GVTO first, followed by the down-sampling and up-sampling GVTOs. 

\subsubsection{Size-preserving GVTO}

% size-preserving GVTO

Given the input third-order tensor $\boldsymbol{\mathscr{X}} \in \mathbb{R}^{d \times h \times w \times c}$ representing $c$ feature maps of the spatial size $d \times h \times w$, the size-preserving GVTO performs three independent $1 \times 1 \times 1$ convolutions on $\boldsymbol{\mathscr{X}}$ and obtains three tensors, namely the query~($\boldsymbol{\mathscr{Q}}$), key~($\boldsymbol{\mathscr{K}}$), and value~($\boldsymbol{\mathscr{V}}$) tensor, where $\boldsymbol{\mathscr{Q}}$, $\boldsymbol{\mathscr{K}}$, $\boldsymbol{\mathscr{V}} \in \mathbb{R}^{d \times h \times w \times c}$. Afterwards, $\boldsymbol{\mathscr{Q}}$, $\boldsymbol{\mathscr{K}}$, $\boldsymbol{\mathscr{V}}$ are unfolded along the channel dimension~\cite{kolda2009tensor} into matrices $\boldsymbol{Q}$, $\boldsymbol{K}$, $\boldsymbol{V} \in \mathbb{R}^{c \times dhw}$. These matrices go through the attention operator defined as
\begin{equation}
\boldsymbol{Y} = \boldsymbol{V} \cdot \textsc{Normalize}(\boldsymbol{K}^T\boldsymbol{Q}) \in \mathbb{R}^{c \times dhw}, \nonumber
\end{equation}
where $\textsc{Normalize}(\cdot)$ is a normalization function that normalizes each column of $\boldsymbol{Q}^T\boldsymbol{K} \in \mathbb{R}^{dhw \times dhw}$. Specifically, the size-preserving GVTO simply uses $1/dhw$ as the normalization function:
\begin{equation}
\boldsymbol{Y} = \boldsymbol{V}\frac{\boldsymbol{K}^T\boldsymbol{Q}}{dhw}  = \frac{1}{dhw} \boldsymbol{V}\boldsymbol{K}^T\boldsymbol{Q} \in \mathbb{R}^{c \times dhw}, \nonumber
\end{equation}
where $dhw$ is the second dimension of $\boldsymbol{Q}$ and subjected to corresponding changes in the down-sampling and up-sampling GVTOs.
After the attention operator, the matrix $\boldsymbol{Y}$ is then folded back to a tensor $\boldsymbol{\mathscr{Y}} \in \mathbb{R}^{d \times h \times w \times c}$. The final outputs of the size-preserving GVTO is the summation of $\boldsymbol{\mathscr{X}}$ and $\boldsymbol{\mathscr{Y}}$, which means a residual connection from the inputs to the outputs~\cite{he2016deep}. In particular, we use the pre-activation technique as well~\cite{he2016identity}. As a result, the size-preserving GVTO preserves the dimension of the inputs (Supplementary Fig. 13e).

\subsubsection{Down-sampling and up-sampling GVTOs}

% other GVTOs

The extension from the size-preserving GVTO to the down-sampling and up-sampling GVTOs is achieved by changing the convolutions that compute $\boldsymbol{\mathscr{Q}}$, $\boldsymbol{\mathscr{K}}$, $\boldsymbol{\mathscr{V}}$. We take the down-sampling GVTO as an example for illustration. Given the same input tensor $\boldsymbol{\mathscr{X}} \in \mathbb{R}^{d \times h \times w \times c}$, we use a $3 \times 3 \times 3$ convolution with stride $2$ to obtain $\boldsymbol{\mathscr{Q}} \in \mathbb{R}^{d/2 \times h/2 \times w/2 \times 2c}$ and two independent $1 \times 1 \times 1$ convolutions to generate $\boldsymbol{\mathscr{K}} \in \mathbb{R}^{d \times h \times w \times 2c}$ and $\boldsymbol{\mathscr{V}} \in \mathbb{R}^{d \times h \times w \times 2c}$. The following computation is the same; that is, $\boldsymbol{\mathscr{Q}}$, $\boldsymbol{\mathscr{K}}$, $\boldsymbol{\mathscr{V}}$ are unfolded along the channel dimension into matrices $\boldsymbol{Q} \in \mathbb{R}^{2c \times dhw/8}$ and $\boldsymbol{K}$, $\boldsymbol{V} \in \mathbb{R}^{2c \times dhw}$, which are fed into the same attention operator and output the matrix $\boldsymbol{Y} \in \mathbb{R}^{2c \times dhw/8}$. Folding it back results in a tensor $\boldsymbol{\mathscr{Y}} \in \mathbb{R}^{d/2 \times h/2 \times w/2 \times 2c}$. Comparing the dimensions of $\boldsymbol{\mathscr{X}}$ and $\boldsymbol{\mathscr{Y}}$, we achieve a down-sampling process that halves the size along each spatial dimension of feature maps but doubles the channel dimension. We complete the down-sampling GVTO by adding the residual connection in two ways, corresponding to two versions of the down-sampling GVTO (Supplementary Fig. 13a-b). One is to perform an extra $3 \times 3 \times 3$ convolution with stride $2$ through the residual connection from $\boldsymbol{\mathscr{X}}$ to $\boldsymbol{\mathscr{Y}}$, in order to transform $\boldsymbol{\mathscr{X}}$ to have the same dimension as $\boldsymbol{\mathscr{Y}}$; the other is to directly add $\boldsymbol{\mathscr{Q}}$ to $\boldsymbol{\mathscr{Y}}$, based on the fact that $\boldsymbol{\mathscr{Q}}$ is obtained from $\boldsymbol{\mathscr{X}}$.

The up-sampling GVTO is dual to the down-sampling GVTO. Instead of using a convolution with stride $2$, it uses a $3 \times 3 \times 3$ transposed convolution with stride $2$ to obtain $\boldsymbol{\mathscr{Q}} \in \mathbb{R}^{2d \times 2h \times 2w \times c/2}$. In addition, the other two $1 \times 1 \times 1$ convolutions generate $\boldsymbol{\mathscr{K}} \in \mathbb{R}^{d \times h \times w \times c/2}$ and $\boldsymbol{\mathscr{V}} \in \mathbb{R}^{d \times h \times w \times c/2}$. The up-sampling GVTO doubles the size along each spatial dimension of feature maps but halves the channel dimension and also has two versions corresponding to different residual connections (Supplementary Fig. 13c-d).

\subsubsection{Advantages of GVTOs}

% advantages

It is noteworthy that, each spatial location in the output tensor of GVTOs has access to all the information in the input tensor, and is able to selectively use or ignore information. We illustrate this point by regarding $\boldsymbol{\mathscr{X}} \in \mathbb{R}^{d \times h \times w \times c}$ as $d \times h \times w$ $c$-dimensional vectors, where each vector represents the information in a spatial location. In this view, each vector has a one-to-one correspondence to each column in $\boldsymbol{K}$ and $\boldsymbol{V}$ in GVTOs, respectively. Revisiting the attention operator, each column in $\boldsymbol{Y}$ is a vector representation of each spatial location in the output tensor, and has a one-to-one correspondence to each column in $\boldsymbol{Q}$. Moreover, each column in $\boldsymbol{Y}$ is computed as the weighted sum of columns in $\boldsymbol{V}$, whose weights are determined by the interaction between the corresponding column in $\boldsymbol{Q}$ and all columns in $\boldsymbol{K}$. The weights can be viewed as filters of the amount of information from each spatial location in the inputs to the outputs. In addition, as both $\boldsymbol{Q}$ and $\boldsymbol{K}$ are computed from the input tensor, the weights are input-dependent. Therefore, GVTOs achieve the dynamic non-local information aggregation.

\subsubsection{Comparisons with Fully-Connected Layers}

It is important to note that the proposed GVTOs are different from fully-connected~(FC) layers in fundamental ways, although they both allow each output unit to use information from the entire input. Compared to FC layers, outputs in GVTOs are computed based on relations among inputs. Thus the weights are input-dependent, rather than learned and fixed during prediction as in FC layers. The only trainable parameters in GVTOs are the convolutions to compute $\boldsymbol{\mathscr{Q}}$, $\boldsymbol{\mathscr{K}}$, $\boldsymbol{\mathscr{V}}$, whose sizes are independent of input and output sizes. As a consequence, GVTOs allow variable-size inputs, and the positional correspondence between inputs and outputs is preserved in GVTOs. In contrast, FC layers require fixed-size inputs and positional correspondence is lost.

\subsection{Training loss}

% loss: 'MSE' (Mean Squared Error) or 'MAE' (Mean Absolute Error)

GVTNets are trained in an end-to-end fashion with two options of the loss functions.
One is the mean squared error~(MSE):
\begin{equation}
\mathscr{L}_{MSE}(y,\hat{y})= \frac{1}{N} \sum_{i=1}^{N} (y_i - \hat{y}_i)^2, \nonumber
\end{equation}
where $y$ represents the ground truth image,  $\hat{y}$ represents the model's predicted image, and $N$ represents the total number of voxels in the image.
The other is the mean absolute error~(MAE):
\begin{equation}
\mathscr{L}_{MAE}(y,\hat{y}) = \frac{1}{N} \sum_{i=1}^{N} |y_i - \hat{y}_i|. \nonumber
\end{equation}

% optimizer: Adam {kingma2015adam}

Both MSE and MAE measure the differences between the predicted image and the ground truth image. The training process applies the Adam optimizer~\cite{kingma2015adam} with a user-chosen learning rate to minimize the loss.

\subsection{Evaluation metrics}

\subsubsection{Pearson correlation coefficient}

Pearson correlation coefficient~($r$) is computed as
\begin{equation}
r(y,\hat{y}) = \frac{\sum_{i=1}^{N} (y_i - \mu_y)(\hat{y}_i - \mu_{\hat{y}})}{\sqrt{\sum_{i=1}^{N} (y_i - \mu_y)^2 \sum_{i=1}^{N} (\hat{y}_i - \mu_{\hat{y}})^2}}, \nonumber
\end{equation}
where $\mu_y$ and $\mu_{\hat{y}}$ are the mean of voxel intensities in $y$ and $\hat{y}$, respectively.

\subsubsection{Normalized root-mean-square error}

The root-mean-square error~(RMSE) is computed as
\begin{equation}
RMSE(y,\hat{y}) = \sqrt{\mathscr{L}_{MSE}(y,\hat{y})}. \nonumber
\end{equation}
The normalized root-mean-square error~(NRMSE) simply adds a normalization function on $y$ and $\hat{y}$, respectively.
In our tools and experiments, we apply the same percentile-based normalization and transformation as in M. Weigert et al.~\cite{weigert2018content}. Concretely, the normalized root mean square error is defined by
\begin{equation}
NRMSE(y,\hat y) = \sqrt{\min_\phi \mathscr{L}_{MSE}(\phi(\hat y), \mathcal{N}(y, 0.1, 99.9))}, \nonumber
\end{equation}
where
\begin{equation}
\mathcal{N}(y, 0.1, 99.9)=\frac{y-percentile(y, 0.1)}{percentile(y,99.9)-percentile(y, 0.1)} \nonumber
\end{equation}
is the percentile-based normalization, and $\phi(\hat y)=\alpha\hat y+\beta$ denotes a transformation that scales and shifts $\hat y$. During the implementation, we let $\alpha =\frac{Cov(y-\bar y, \hat y - \bar{\hat y})}{Var(\hat y-\bar{\hat y})}$ and $\beta=0$ to obtain $\phi(\hat y)$ so that the MSE is minimized.

\subsubsection{Structural similarity index}

The structural similarity index~(SSIM)~\cite{wang2004image} is computed as
\begin{equation}
SSIM(y,\hat{y}) = \frac{(2\mu_y\mu_{\hat{y}} + c_1)(2\sigma_{y\hat{y}} + c_2)}{(\mu_y^2 + \mu_{\hat{y}}^2 + c_1)(\sigma_y^2 + \sigma_{\hat{y}}^2 + c_2)}, \nonumber
\end{equation}
where $\sigma_y$ is the variance of $y$, $\sigma_{\hat{y}}$ is the variance of $\hat{y}$, $\sigma_{y\hat{y}}$ is the covariance of $y$ and $\hat{y}$, and $c1=(0.01L)^2$, $c2=(0.03L)^2$ are two constant parameters of SSIM. Here, $L$ represents the range of intensity values and is set to $1$.

\subsection{Task-specific configurations}

The settings of our device are - GPU: Nvidia GeForce RTX 2080 Ti 11GB; CPU: Intel Xeon Silver 4116 2.10GHz; OS: Ubuntu 16.04.3 LTS.

\subsubsection{Label-free prediction of 3D fluorescence images from transmitted-light microscopy}

The basic GVTNet used in the experiments of label-free prediction of 3D fluorescence images is illustrated in Supplementary Fig. 2. The network has depth 4, where the skip-connections add feature maps from the encoder to the decoder. In particular, the bottom block of the basic GVTNet is the size-preserving GVTO (Supplementary Fig. 11e). The number of feature maps after the initial convolution is set to 32. Batch normalization~\cite{ioffe2015batch} with the momentum of 0.997 and epsilon of 0.00001 is applied before each ReLU activation function.

The 13 subtasks corresponding 13 different subcellular structures are performed separately and independently. To train the GVTNet, the 30 pairs of training images are randomly cropped into patches of size $64\times64\times32$ and each training batch contains 16 pairs of patches. We minimize the MSE loss using the Adam optimizer with a learning rate of 0.001 for 70,000 to 100,000 minibatch iterations, depending on different subtasks. The training procedure lasts approximately 11h15m to 15h45m for each of the 13 datasets~\cite{ounkomol2018label}.

\subsubsection{Context-aware 3D image denoising}

The GVTNet used in the image denoising tasks is illustrated in Supplementary Fig. 4. It follows a 3D U-Net framework of depth 3, i.e., including 2 down-sampling and up-sampling operators, respectively. The skip-connections merge feature maps from the encoder to the decoder by concatenation instead of addition. The bottom block is the size-preserving GVTO and two up-sampling operators are the up-sampling GVTOs v2 (Supplementary Fig. 11d). The number of feature maps after the initial convolution is set to 32. No batch normalization is applied.

We use the MAE loss with the Bayesian deep learning technique~\cite{kendall2017uncertainties} (Supplementary Note 1) to train the GVTNet. The training patch size is $64\times64\times16$. We train the model with a batch size of 16 and a base learning rate of 0.0004 with a decay rate 0.7 for every 10,000 minibatch iterations. The training procedure takes 50 epochs and lasts about 5h45m and 4h50m for the Planaria and Tribolium datasets~\cite{weigert2018content}, respectively.

\subsubsection{Content-aware 3D to 2D image projection}

The model for surface projection is composed of a 3D to 2D projection network and a 2D denoising network, as illustrated in Supplementary Fig. 10. The projection network predicts the probability of each voxel in the 3D input image belonging to the 2D surface, and uses summation weighted by the predicted probabilities along the Z-axis to finish the projection. The probabilities are estimated by a GVTO-augmented CNN. The following 2D denoising network is simply a 2D version of the GVTNet used in the image denoising tasks.

During training, the 3D input patch size is $64\times64\times50$ and the 2D ground truth patch size is $64\times64\times1$. The other training settings are the same as those in image denoising experiments, except that we do not use the Bayesian deep learning technique. The training procedure lasts 4h55m for the Flywing dataset~\cite{weigert2018content}.

\subsection*{Reporting summary}

Further information on research design can be found in the Nature Research Reporting Summary linked to this article.

\section*{Data availability}

Datasets for label-free prediction of 3D fluorescence images from transmitted-light microscopy~\cite{ounkomol2018label} can be downloaded from \url{https://downloads.allencell.org/publication-data/label-free-prediction/index.html}.
Datasets for context-aware 3D image denoising and 3D to 2D image projection~\cite{weigert2018content} can be downloaded from \url{https://publications.mpi-cbg.de/publications-sites/7207}.

\section*{Code availability}

The code for GVTNets training, prediction and evaluation (in Python/TensorFlow) is publicly available at \url{https://github.com/divelab/GVTNets} and \url{http://doi.org/10.5281/zenodo.4285769}.

% \bibliography{reference}

\end{document}